# Thermoelectric Performance Boost by Chemical Order in Epitaxial L2$_1$ (100) and (110) Oriented undoped Fe$_2$VAl Thin Films: An Experimental and Theoretical Study

José María Domínguez-Vázquez[a], Olga Caballero-Calero[a], Ketan Lohani[a], José J. Plata[b], Antonio M. Marquez[b], Cristina V. Manzano[a], Miguel Ángel Tenaguillo[a], Hiromichi Ohta[c], Alfonso Cebollada[a], Andres Conca[a,*], and Marisol Martín-González[a].

This study demonstrates the direct correlation between the presence of the L2$_1$ ordered phase and the large enhancement in the thermoelectric performance of Fe$_2$VAl thin films deposited on MgO and Al$_2$O$_3$ substrates at temperatures varying between room temperature and 950 °C. We employ both experimental techniques and computational modeling to analyze the influence of crystallographic orientation and deposition temperature on the thermoelectric properties, including the Seebeck coefficient, electrical conductivity, and thermal conductivity. Our findings indicate that the presence of the L2$_1$ phase significantly enhances the power factor (PF) and figure of merit (zT), surpassing previously reported values for both bulk and thin film forms of Fe$_2$VAl, achieving a PF of 480 µW/m·K$^2$ and a zT of 0.025.

## Broader context

Thermoelectric materials transform waste heat into valuable electricity, addressing a critical inefficiency in global energy systems where over 70% of primary energy is lost as heat. Developing high-performance, non-toxic thermoelectric materials from earth-abundant elements represents a sustainable pathway to harness this wasted energy resource. Fe$_2$VAl Heusler alloys offer a promising alternative to conventional thermoelectric materials that contain scarce or toxic elements, potentially enabling widespread deployment in industrial waste heat recovery, IoT device powering, and wearable technology applications. Our demonstration of significantly enhanced thermoelectric performance through chemical ordering optimization provides a novel approach to developing eco-friendly thermoelectrics without relying on expensive or environmentally problematic elements. This work contributes to the broader goal of creating energy-efficient systems that reduce primary energy consumption while meeting growing global electricity demands through sustainable materials engineering.

[a.] Instituto de Micro y Nanotecnología, IMN-CNM, CSIC (CEI UAM+CSIC), Isaac Newton 8, E-28760 Tres Cantos, Madrid, Spain
[b.] Dpto de Química Física, Facultad de Química, Universidad de Sevilla, Sevilla (Spain).
[c.] Research Institute for Electronic Science, Hokkaido University, N20W10, Kita, Sapporo, 001–0020, Japan.
*Corresponding author: andres.conca@csic.es
† Supplementary Information available

## Introduction

Thermoelectric materials have the unique ability to convert thermal gradients into electrical voltage, enabling direct energy conversion between heat and electricity. This property positions them as key players in addressing the global energy crisis, particularly in waste heat recovery and energy-efficient cooling applications across various industrial sectors. By capturing wasted thermal energy, they can significantly improve the overall efficiency of energy systems. Additionally, these materials offer a sustainable, low-maintenance solution for powering Internet of Things (IoT) devices, wearable technologies, and biomedical sensors. The efficiency of thermoelectric energy conversion is governed by the dimensionless figure of merit, $zT$, which is defined as $zT=S^2 \cdot \sigma \cdot T/\kappa$, being $S$ the Seebeck coefficient, $\sigma$ the electrical conductivity, $T$ the temperature, and $\kappa$ the thermal conductivity. Thus, high-performing thermoelectric materials are those that exhibit a high $S$ and $\sigma$, while maintaining a low $\kappa$. Bismuth telluride (Bi$_2$Te$_3$)[1,2], lead telluride (PbTe)[3,4] or Selenides like SnSe[5,6], Ag$_2$Se[7] or Cu$_2$Se[8] have been extensively studied as the best-performing thermoelectric materials for different temperature applications. However, challenges related to cost, toxicity, and the scarcity of some of these materials limit their practical application.

In recent years, Heusler alloys —specifically, half-Heusler XYZ[9–11] and full-Heusler X$_2$YZ[12–14] compounds— have emerged as promising alternatives due to their tunable semiconducting properties and earth-abundant, eco-friendly compositions. Notably, Fe$_2$VAl[15–18], a full-Heusler alloy, has attracted significant attention for its potential thermoelectric applications. Fe$_2$VAl exhibits relatively high-power factors (PF =

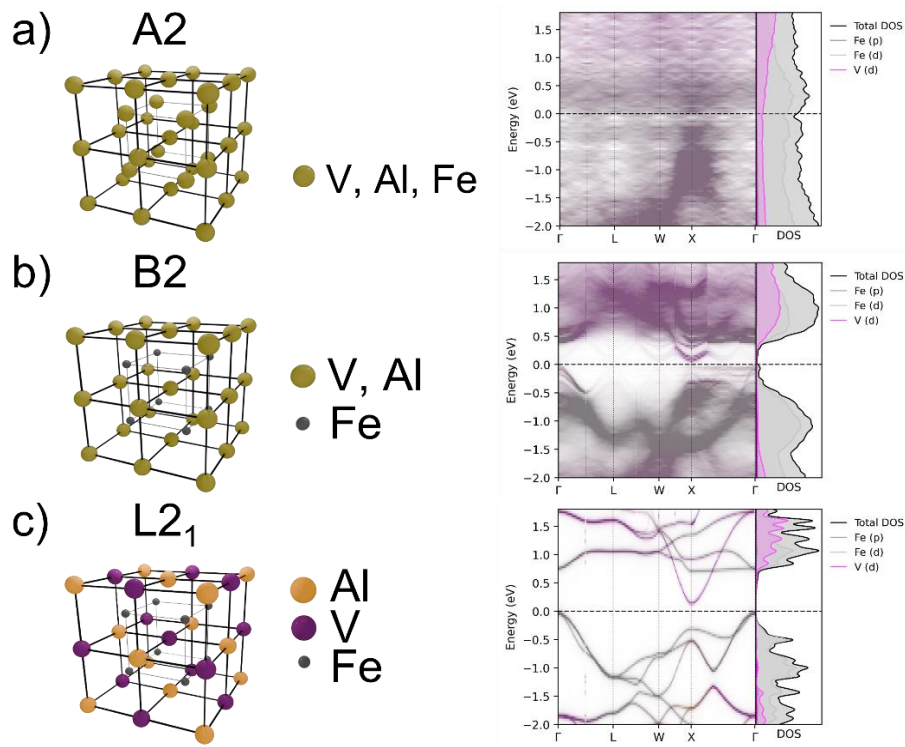

Figure 1: *Crystalline structures of Fe₂VAl for different degrees of chemical order: A2 [a)], B2 [b)] and L2$_1$ [c)], a scheme of the structure is shown along with its band structure and DOS.*

$S^2 \cdot \sigma$) and can be tailored to p- or n-type conductivity through stoichiometric variations or doping with abundant elements, such as Ti and Cr[19,20] for p-type and W, Si, or Ta[21–24] for n-type.

Recent breakthrough research[25] has demonstrated that Fe₂VAl exhibits exceptional thermoelectric properties at the Anderson transition, a quantum phase transition from localized to mobile electron states, where conditions for ideal thermoelectric performance are met. Complementary to this, other works have proved that electronic band engineering[26], off-stoichiometric doping[17] or self-substitution[27,28] can enhance significantly the power factor (PF=$S^2 \cdot \sigma$) and the figure of merit (zT). Our approach focusses on the chemical order of the phase, representing a distinct yet complementary strategy to these advances.

Fe₂VAl possesses a cubic structure with varying degrees of chemical order that strongly influence its electronic band structure and thermoelectric properties. These different structures are illustrated in the left column of *Figure 1*, where the chemically disordered cubic phase, A2, which has all three elements occupying lattice positions at random; the B2 phase, where Al and V occupy positions indistinctively; and finally, the fully ordered L2$_1$ phase, where each element atom occupy univocally specific lattice sites, are schematically depicted. These three phases, characterized by increasing chemical order, endorse the material with specific band structure features and as a consequence, different thermoelectric properties. This is shown in the right column of *Figure 1* where theoretically calculated band structures and DOS are shown for A2, B2 and L2$_1$ phases, respectively. As can be seen in the figure, chemical order causes the existence of a defined gap in the band structure, which eventually enhances the Seebeck coefficient, being this Seebeck enhancement particularly large when the L2$_1$ structure is attained. Although other properties such as electrical or thermal conductivity seem to evolve towards worsening thermoelectric efficiency with greater chemical ordering, the presented calculations predict that this increase in the Seebeck coefficient makes L2$_1$ structure the one that reaches the highest value of *zT*.

Conceptually, preparing thermoelectric materials in thin film form offer numerous advantages over bulk materials due to reduced thermal conductivity at the nanoscale, their applicability as covering functional layers or their use of small amounts of material. Fe₂VAl thin films, in particular, have demonstrated enhanced thermoelectric performance, especially when grown via sputtering—a scalable, industry-friendly deposition method—, although most studies focus on polycrystalline films. However, while epitaxial films, with long range crystalline coherence, provide an ideal platform for studying the intrinsic thermoelectric response of model systems and allow for more accurate comparisons with theoretical predictions, investigations into epitaxial or single-oriented thin films and their thermoelectric properties remain scarce.

In this work, we govern the appearance of the previously mentioned cubic phases in Heusler alloy Fe₂VAl (100) and (110) oriented thin films grown by sputtering from stoichiometric targets and study experimentally their corresponding thermoelectric properties. Chemical order and crystalline orientation are simply controlled by adequate selection of deposition temperature, $T_{dep}$, and substrate orientation. In our case MgO (100) and Al₂O₃ (11-20) substrates promote the growth in the (100) and (110) directions, respectively, and allow us to obtain the desired cubic phase with controllable degree of chemical order at will. Specifically, we demonstrate the relevance of growing at high $T_{dep}$ in promoting the $L2_1$ phase, for which our results show a notable enhancement in PF and $zT$ values compared to previously reported $Fe_2VAl$ thin films, underscoring the potential of these materials for practical thermoelectric applications.

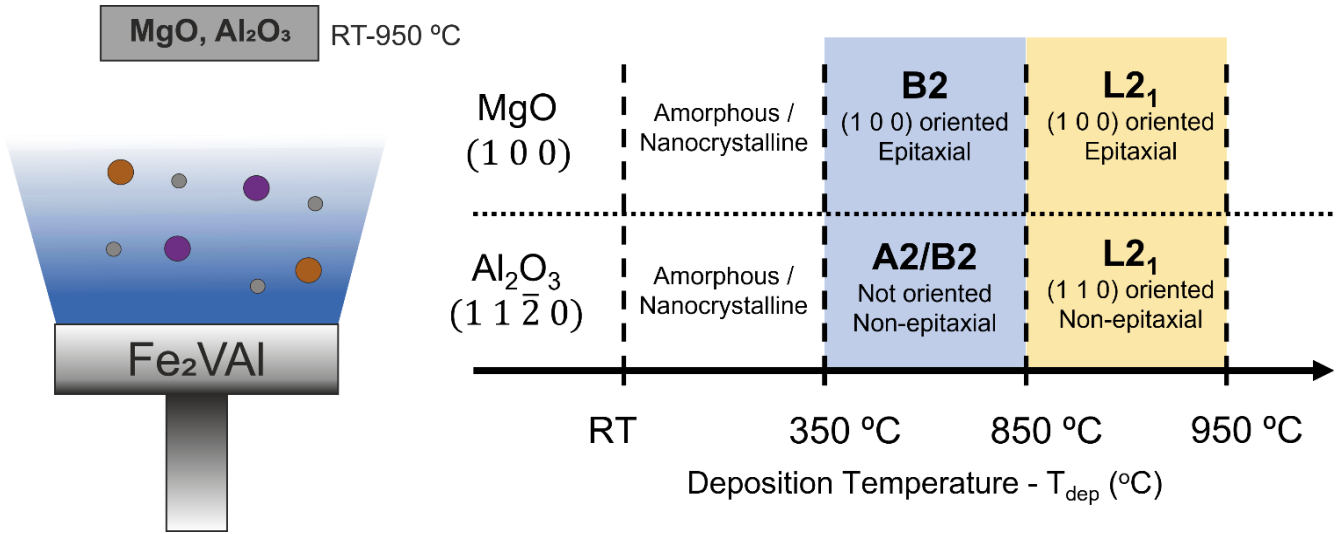

Figure 2: Highest chemically ordered crystalline structures and type of growth obtained for ranges of deposition temperature ($T_{dep}$) between RT and 950°C for the two different substrates used in this work (MgO and $Al_2O_3$). Two main ranges are highlighted: in blue, 350-850 °C, with B2 or A2 as the highest chemically ordered structure present, and in yellow 850-950°C, where $L2_1$ presence is observed.

## Growth and characterization of $Fe_2VAl$ thin films

A summary of the crystalline phases determined by X-ray diffraction measurements for the studied films is shown in *Figure 2* (full details in SI). Generally speaking, for both substrates, increasing $T_{dep}$ leads to a higher degree of crystallinity. Films grown below 350 °C are nanocrystalline or of amorphous in nature, as no diffraction peaks are observed for them. For $T_{dep}$ between 350 and below 850 °C, the highest ordered structure present in the films is B2 for films deposited over MgO and, either B2 or A2 for films deposited over $Al_2O_3$ (depending on $T_{dep}$). In this range of $T_{dep}$ films grown on MgO are epitaxial, (100) oriented, due to the favourable cube-on-cube symmetry matching, while those grown on $Al_2O_3$ are polycrystalline. Finally, for $T_{dep}$ higher than or equal to 850 °C in both substrates, the $L2_1$ phase is detected. In this last range of $T_{dep}$, films grown over MgO are epitaxial but, due to the lower symmetry of the $Al_2O_3$ (11$\bar{2}$0) face, films deposited on this substrate are only (110) textured or single-oriented, but not epitaxial, lacking in-plane crystalline coherence. The complete X-ray diffraction (XRD) data and characterization can be found in the supplementary information (SI).

As it is seen in *Figure 1*, theoretical calculations predict strong effects of the chemical order on both band structure and thermoelectric properties of the compound. To experimentally confirm this correlation, in *Figure 3a* we show the $T_{dep}$ dependence of the normalized intensity of the (1 1 1) XRD peak, proving $L2_1$ presence in the film, along with the measured Seebeck coefficient, conductivity and Power Factor (PF). On the one hand, a twofold increase in the Seebeck coefficient (positive for all samples, and therefore p-type) is observed as $T_{dep}$ rises between 350 and 800 °C, with quite a linear behaviour. For 850 °C and above, and along with a gradual increase of the (111) reflection in the XRD analysis, and therefore of the presence of $L2_1$ phase, the Seebeck coefficient experiences another twofold increase.

On the other hand, the electrical conductivity experiences a continuous gradual decrease as $T_{dep}$ increases from room temperature to 950 °C, probably due to the rougher, more pronouncedly agglomerated morphology in the films observed via AFM and SEM (discussed later), which may enhance charge carrier scattering at higher $T_{dep}$. Interestingly, this decrease saturates when the $L2_1$ phase is obtained, and therefore, band structure effects on the electrical conductivity associated with the presence of this ordered phase cannot be discarded.

Combining the Seebeck coefficient and electrical conductivity measurements, we observe two distinct regions in the Power Factor dependence on deposition temperature ($T_{dep}$). In the absence of the $L2_1$ phase (lower $T_{dep}$): The PF shows a linear, gradual increase with $T_{dep}$. While, the $L2_1$ phase is present (higher $T_{dep}$) an abrupt change in the slope of PF vs $T_{dep}$, with a steeper increase. While minor variations exist between substrates, the overall trend of increasing PF with higher $T_{dep}$ is consistent for both MgO and $Al_2O_3$ substrates. The similarity in PF values between films grown in different crystalline orientations ((100) on MgO and (110) on $Al_2O_3$) suggests isotropic electronic transport properties in $Fe_2VAl$. This isotropy is further confirmed by in-plane rotation measurements, where the Seebeck coefficient remains constant regardless of the thermal gradient direction relative to the crystalline axis.

Thermal conductivity (κ) measurements by TDTR at room temperature for films deposited on MgO yield to 4.4 ± 0.2 W/m·K for the $B_2$ phase ($T_{dep}$ = 350°C) and 4.6 ± 0.2 W/m·K for the $L2_1$ phase ($T_{dep}$ = 950°C). Interestingly, these κ values are within the experimental error of each other, indicating that the structural ordering primarily affects the power factor rather than the thermal conductivity. Given that $Fe_2VAl$ is a cubic phase and theoretically isotropic, we can calculate the figure of merit (zT) for both phases: B₂ phase: zT = 0.004 and L2₁ phase: zT = 0.016. This represents a fourfold increase in zT for the L2₁ phase compared to the B₂ phase, highlighting the significant impact of structural ordering on the thermoelectric performance (mainly in the power factor) of Fe₂VAl thin films.

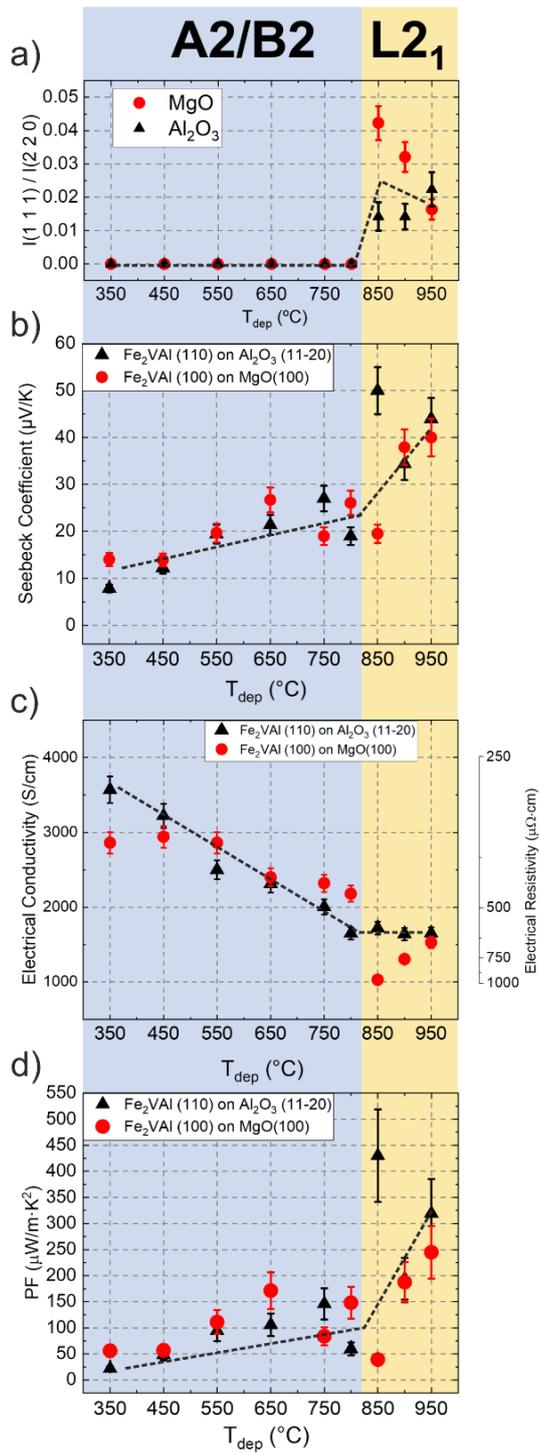

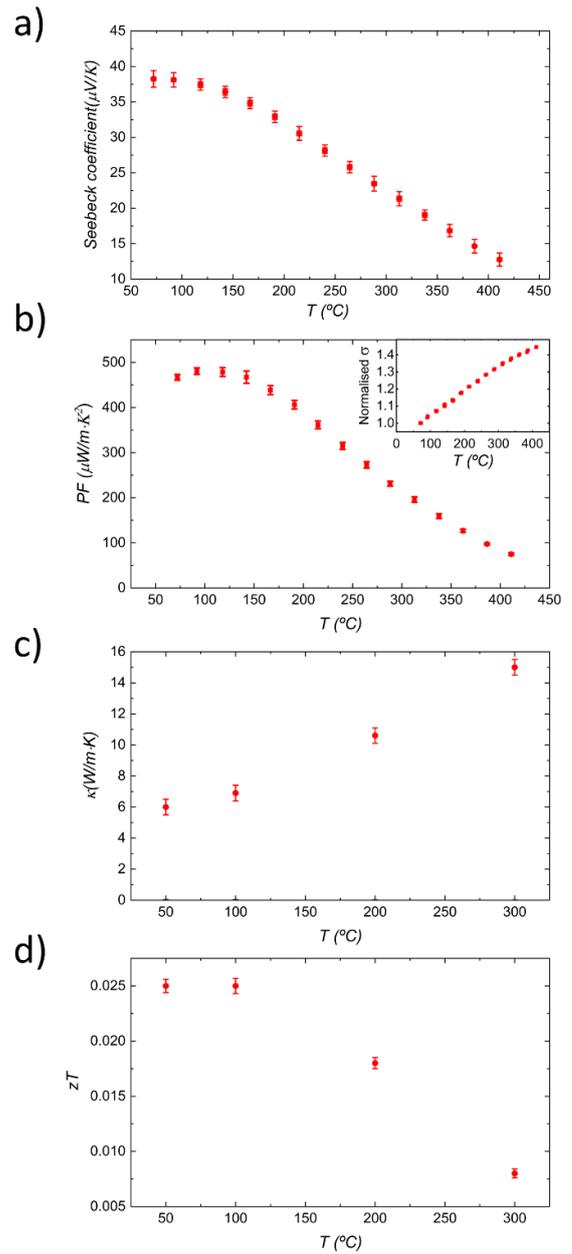

Figure 3: *Temperature dependence of Seebeck coefficient a), power factor b), thermal conductivity (κ) c) and zT d) of a Fe₂VAl thin film deposited at 900°C. An inset was included to show the evolution with temperature of electrical conductivity normalized.*

Figure 4: *Thermoelectric properties and L2₁ presence as a function of deposition temperature (T_dep) of Fe₂VAl thin films. a) relative intensity of the (1 1 1) x-Ray diffraction peak with respect to the (2 2 0) peak, b) Seebeck coefficient, c) Conductivity and d) PF were plotted highlighting the samples with L2₁ presence in yellow, and the ones with B2 or A2 structure in blue. Along with data points a guide to the eye of the evolution of the distinct properties was included (dashed line).*

*Figure 4* shows the evolution with temperature of the Seebeck coefficient, PF, normalised σ, κ and zT of a Fe₂VAl thin film with L2₁ order (deposited at 900 °C over MgO). The plot shows a gradual decrease of the Seebeck coefficient with increasing temperature and, therefore, a decrease in the power factor. A complementary increase in the electrical and thermal conductivity is observed. The inset included in *Figure 4* b), showing the temperature dependence of normalised σ, shows a semiconductor-like behaviour, yielding an increasing σ with rising temperature. The range of temperatures where the maximum of PF and zT is obtained is between 50 and 150 °C, which agrees with previous works [17,25]. The PF peaks at a value of 480 μW/m·K² which corresponds to the largest zT value of 0.025, stable in the mentioned temperature range. The absolute maximal values of zT, together with the value measured at room temperature, are significant since they compare in a very positive manner with the reported in the literature. For a best visualization of this fact, we next make a quantitative comparison between our highest obtained PF values (measured at RT) with those reported in the literature (e.g., [17,18,21,29–33]) for both thin film and bulk materials. This is shown in *Figure 5 a)*, where our obtained PF values are more than twofold of those obtained for other works. The work by Furuta *et al*.[32] is of particular relevance, as they also reported epitaxial growth of $Fe_2VAl$ films, albeit on $MgAl_2O_4$ substrates, which offer a smaller lattice mismatch compared to MgO. However, despite this, their PF values were lower, likely due to lower electrical conductivity and Seebeck coefficient. Furuta's films, with a thickness of 1 μm, were significantly thicker than those in this study and exhibited a weaker (111) diffraction peak, suggesting a lower proportion of the $L2_1$ phase. Their use of lower deposition temperatures (up to 500 °C) likely limited the chemical ordering and $L2_1$ phase development, resulting in reduced thermoelectric performance. In contrast, our study demonstrates the importance of high $T_{dep}$ for achieving $L2_1$ phase and maximizing thermoelectric performance. This finding is reinforced in *Figure 5 b)*, where the thermoelectric figure of merit (*zT*), measured at RT, of $Fe_2VAl$ films deposited over MgO at 350 °C and 950 °C are compared with values found in the literature ([18,30,32–34]). A four-fold increase in the zT value can be seen between these two samples of this work, being the one deposited at a lower temperature (where the $L2_1$ phase is absent) in the range of the values found in the literature, while the one deposited at the highest temperature surpasses these values in more than a twofold increase. In *Figure 5 c)* the maximum zT (obtained at 50-100 °C) of 0.025 is compared with the maximum zT in temperature obtained in [18,30,32–34], where our reported maximum value surpasses by an almost four-fold increase the highest value found in the literature. Although the values of *Furuta et al.* and *Yamada et al.* were measured at room temperature and do not report values at higher temperatures.

The *zT* enhancement is caused by the previously mentioned increase in PF, as the thermal conductivity was found to be almost the same for both films (4.4±0.2 and 4.6±0.2 W/m·K for $T_{dep}$ of 350 °C and 950 °C respectively). This result highlights the importance of $L2_1$ ordering for improving the thermoelectric efficiency and the subsequential relevance of $T_{dep}$ as a growth parameter of $Fe_2VAl$ films. The $T_{dep}$ does not only influence the nature of the obtained crystalline order (amorphous, textured or epitaxial) and the final chemical order of the structure (A2, B2 or $L2_1$). It affects the growth process resulting in a change of the crystallite and grain sizes, and results in a different film morphology. A different distribution and sizes of crystalline grains may affect the thermoelectric properties, especially thermal conductivity, and therefore, a study of the morphology on the films was made. In *Figure 6*, SEM micrographs can be seen for three representative $T_{dep}$ and both substrate cases.

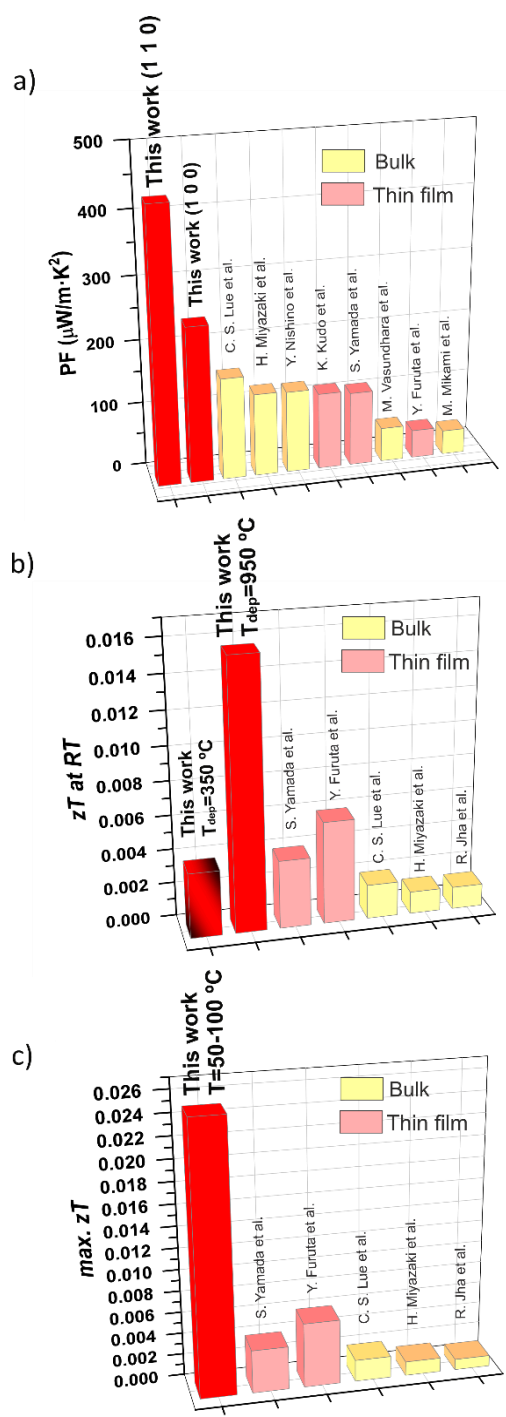

Figure 5: *Thermoelectric properties of $Fe_2VAl$ films of this work compared with bulk and thin film values of the literature. In a) the best values of PF obtained for each orientation are plotted with values obtained from[17,18,21,29–33], in b), zT values at room temperature of $Fe_2VAl$ deposited at 350 °C and 950 °C are plotted along values obtained from[18,30,32–34] and our maximum value obtained at 50-100 °C.*

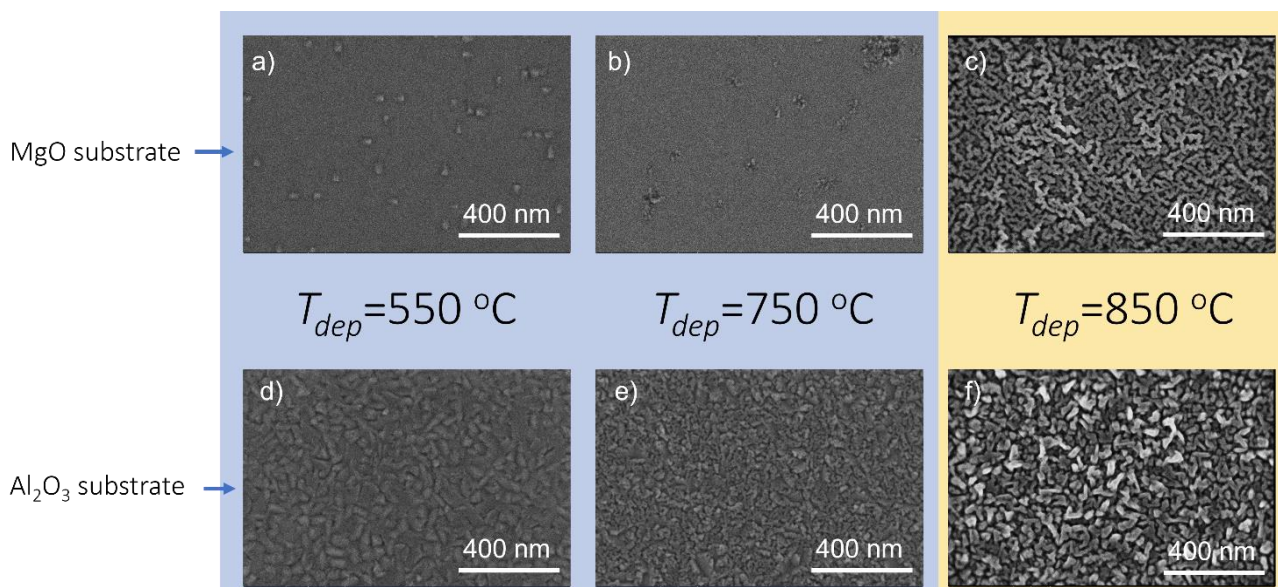

Figure 6: *SEM micrographs of the Fe$_2$VAl thin films grown on MgO and Al$_2$O$_3$. a-c) Films on MgO at 550°C, 750°C, and 850°C, respectively. (d-f) Films on Al$_2$O$_3$ at the same temperatures. Highlighted in blue, the films with B2/A2 as highest chemically ordered structure, while in yellow, the ones with L2$_1$ structure present.*

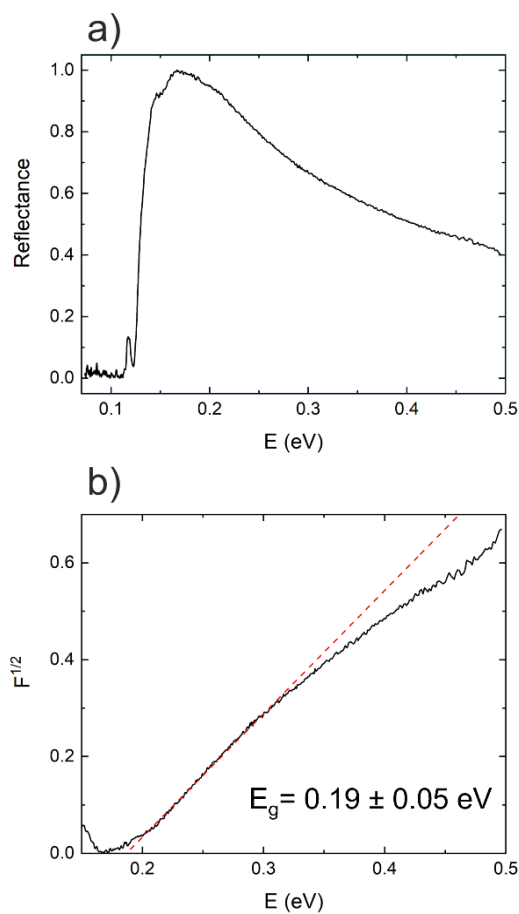

Figure 7: *a) Reflectance versus photon energy for a 155 nm thick Fe$_2$VAl sample deposited at 900°C on a MgO (100) substrate. b) Pankove plot used for the estimation of the band gap (Tauc method was also used, resulting in the same result E$_g$=0.19 ± 0.05 eV).*

The morphology of Fe$_2$VAl films changes drastically with $T_{dep}$ on both substrates, and the roughness increases also with $T_{dep}$. Line profile plots, RMS roughness values and 3D AFM images of these films are shown on the Supporting Information. For both series of films, a change from a planar to a more granular morphology is observed as $T_{dep}$ increases. However, differences between the series also appear, due to the different induced grain formation owed to the distinct growth orientation and substrate-film interaction. The enhanced thermoelectric performance observed in our L2$_1$ ordered Fe$_2$VAl films presents interesting parallels with the work by Garmroudi *et al.*[25] regarding the Anderson transition in similar materials. While the work by Garmroudi *et al.* demonstrates optimal thermoelectric conditions when conductive electrons have approximately the same energy at this quantum phase transition, our approach achieves enhancement through structural ordering that modifies the electronic band structure. The thermal conductivity values we report (4.4±0.2 and 4.6±0.2 W/m·K at room temperature for the B2 and L2$_1$ phases respetively) remain relatively high compared to state-of-the-art thermoelectric materials, suggesting that further optimization through nanostructuring or selective doping could potentially reduce thermal conductivity while preserving the beneficial electronic properties of the L2$_1$ phase. From a practical device implementation perspective, our epitaxial thin film approach offers advantages for integration into microelectronic systems where controlled interfaces and structural coherence are critical for reliable performance in energy harvesting applications.

## Analysis of thermoelectric behaviour: reconciling theory and experiment

As discussed earlier, the chemical order significantly impacts the band structure of the material, thereby influencing its thermoelectric performance. Key features such as the band gap, band curvature, and density of states are crucial for understanding the measured transport properties. For this reason, it is highly interesting to evaluate the possibility of measuring the energy band gap of this alloy. The estimation of the band gap was done by optical methods; therefore, an optically thick (155 nm) sample was deposited at 900 °C over MgO. In *Figure 7* the reflectance spectrum of this sample is plotted along with a Pankove plot (Tauc plot was also used and obtained the same result). The estimation of the band gap of the sample, as derived from optical measurements, reveals a value of 0.19 ± 0. 05 eV. This estimation aligns closely with previous experimental findings, such as those reported by Rai *et al.*[35], who determined a band gap of approximately 0.2 eV. However, it is slightly lower than the value of 0.27 eV reported by Lue *et al.*[36], who utilized NMR data in their analysis. The discrepancies in band gap values across different studies can often be attributed to variations in experimental methodologies and sample conditions, including factors such as film thickness, substrate choice, deposition temperature or crystallinity, which can significantly influence optical properties.

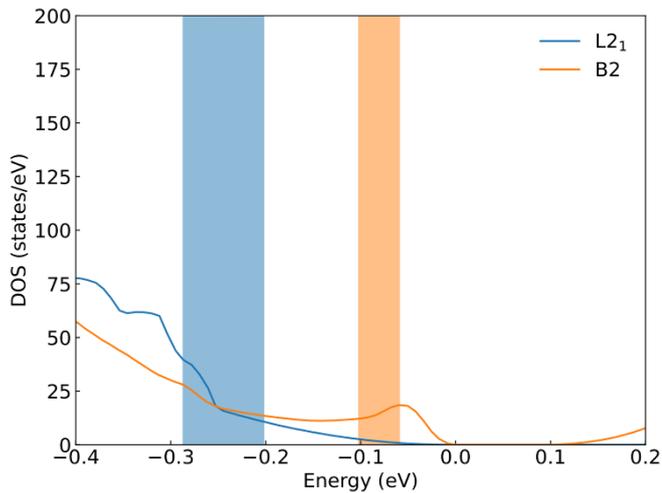

Figure 8: a) Comparison of the DOS valence band edge for L2$_1$ (blue) and B2 (orange) phases. The colored range shows the Fermi energy position based on experimental determined carrier concentration.

The prediction of the electronic structure of Fe$_2$VAl can be challenging and controversial as its experimental characterization. The choice of exchange correlation functional can significantly influence the description of narrow band gap semiconductors when using Density Functional Theory. For instance, General Gradient Approximation (GGA) functionals widely used like PBE [37] or PBE+U[22] have described L2$_1$ Fe$_2$VAl as a semimetal. However, meta-GGA functionals like mBJ, which typically improve the accuracy of GGAs by accounting for the local kinetic energy density, have characterized this material as a narrow band gap semiconductor[38]. Interestingly, these findings complement the work by Garmroudi *et al.*[25] on the quantum phase transition concept in Fe$_2$VAl, where the electronic structure near the Anderson transition creates ideal conditions for thermoelectric performance. Our observation of enhanced Seebeck coefficient in the L21 phase can be understood as a manifestation of optimized electronic structure that maximizes the power factor despite modest decreases in electrical conductivity and no change in the thermal conductivity. This mechanism of band structure engineering through chemical ordering represents a broadly applicable approach that could be exploited in other Heusler systems to enhance thermoelectric performance. In this work, the r2SCAN functional has yielded a band gap of 0.18 eV for the L2$_1$ ordered phase, which aligns well with the band gaps obtained using the mBJ functional and the experimental evidence reported in this work and previous results reported by Garmroudi *et al*[25].

Tackling the modelling of disorder has led to seemingly contradictory results. While Garmroudi *et al*.[25] experimentally found that disorder reduces the band gap and leads to a metallic band structure for A2[25], computational studies have suggested that disorder can increase the band gap[39]. In this work, different degrees of disorder were investigated to monitor the evolution of the band structure and its gap. For low levels of disorder based on V/Al antisite defects, a statistical approach was employed, wherein the band gap of the system was calculated as an ensemble of all possible microstructures. Conversely, B2 and A2 structures were modelled using the Special Quasirandom Structures (SQS) methodology. The calculated band gaps reconcile the previous results. At low concentrations of V/Al antisites, a widening of the band gap is observed, with values of 0.19 and 0.25 eV for 3.125 and 6.25% disorder, respectively. Indeed, the band gap of 0.19 eV at 3.125% of V/Al antisites is very close to the experimentally values obtained in this work. However, at higher levels of disorder, the SQS structures used to model the B2 and A2 phases exhibit a reduction in the band gap (*Figure 1*), ultimately leading to a metallic structure for the A2 phase.

Obtaining a comprehensive understanding of the thermoelectric performance requires more than just a detailed analysis of the band gap. Other key features, such as band curvature and density of states, are also essential to obtain a complete picture of the thermoelectric behavior. The thermal and transport properties of the L2$_1$ phase have been calculated, taking into account the carrier concentration, grain size, and temperature of the synthesized samples (see **Supporting Information**). While the simulated lattice thermal conductivity (3 and 15 W/m·K for films with crystallite sizes of 12 and 41 nm respectively) is in good agreement with experimental results (2.3 and 3.4 W/m·K for films deposited over MgO at $T_{dep}$ of 350 and 950 °C respectively) when grain boundaries are considered in the calculations, the Seebeck coefficient (~90μV/K) and electrical conductivity (~2200 S/cm) are slightly overestimated. It is important to note that effects such as epitaxial strain or texture are not included in the model and may play a role in the overall thermoelectric performance.

To understand the behavior of the transport and thermoelectric properties based on the level of disorder, the band structure and density of states of the three phases are compared. A reduction of the band gap when disorder is increased is in good agreement with the enhancement of the electrical conductivity when the deposition temperature is decreased. In the epitaxial films grown on MgO, where the lattice mismatch is smaller, there is a non-monotonous behavior at high $T_{dep}$ that could be explained by the enlargement of the band gap at small concentrations of V/Al antisites. To explain the large enhancement of the Seebeck coefficient, the slope of the density of states at the edge of the valence band is analyzed for the different levels of disorder. The metallization process due to disorder not only reduces the band gap but also changes the topology of the density of states. The larger the disorder, the flatter the band becomes. In *Figure 8*, the edge of the valence band of the L2$_1$ and B2 models are plotted, highlighting the position of the Fermi level based on the experimentally measured carrier concentrations for MgO and Al$_2$O$_3$ substrates. The larger slope of the DOS at the L2$_1$ phase is in good agreement with the larger Seebeck coefficient obtained experimentally.

It is important to note that a similar behavior of the band structure with chemical disorder has been calculated by Graf *et al.*[40] on the half Heusler of TiNiSn. In their work the increasing degree of disorder results also in a progressive band-gap closure until a metallic band structure is obtained. Therefore, chemical order can play a significant role in the thermoelectric performance for several half and full Heusler compounds.

## Conclusions

This study provides strong evidence that L2$_1$ ordered Fe$_2$VAl thin films exhibit superior thermoelectric performance. Thin films grown with $T_{dep}$ exceeding 850 °C result in higher Seebeck coefficients and, consequently, an enhanced power factor, with the highest values being 430 ± 80 µW/m·K² and 245 ± 50 µW/m·K² for films on Al$_2$O$_3$ and MgO substrates, respectively. A thermal conductivity of 4.6 ± 0.2 W/m·K at room temperature was measured by TDTR means for samples deposited over MgO. The highest zT value of 0.016 was achieved for films deposited at 950 °C, representing a twofold increase compared to the literature.

The drastic changes in the thermoelectric performance of Fe$_2$VAl are closely linked to its electronic structure and the degree of chemical order. Disorder leads to a semiconductor-to-metal transition, which significantly impacts the transport properties. While the larger band gap of the L2$_1$ phase reduces its electrical conductivity, this effect is compensated by the higher Seebeck coefficient. The aforementioned twofold increase in the Seebeck coefficient can be explained by the flattening of the density of states bands during the transition from the L2$_1$ to the B2 phase.

These findings highlight the potential of Fe$_2$VAl as a promising, earth-abundant, non-toxic thermoelectric material that offers significant environmental advantages over conventional materials and pave the way for further optimization through doping with elements like Ti, W, and Ta, as well as investigate approaches to reduce thermal conductivity through nano structuring while preserving the beneficial electronic properties of the L21 phase. The scalable sputtering deposition method demonstrated in this work offers a pathway toward practical implementation in energy harvesting devices for Internet of Things applications, wearable technologies, and industrial waste heat recovery systems, contributing to global energy efficiency goals and climate impact mitigation through reduced primary energy consumption.

## Experimental details

Two series of 80 nm thick Fe$_2$VAl thin films were deposited on MgO (1 0 0) and Al$_2$O$_3$ (1 1 -2 0) substrates at $T_{dep}$ ranging from RT to 950 °C. The deposition was carried out in a UHV chamber (base pressure ~10$^{-9}$ mbar) utilizing DC magnetron sputtering. A stoichiometric commercial Fe$_2$VAl target (Mateck GmbH) was sputtered at 40W and 2.5·10$^{-3}$ mbar Ar pressure, yielding a deposition rate of 1 nm/min. $T_{dep}$ was measured in situ using a calibrated thermocouple located in the sample holder. The structure was characterized by X-ray Diffraction (XRD) measurements performed on a Bruker D8 Discover four-circle diffractometer with a microfocus X-ray source (IµS) (Cu K$_{\alpha}$1) and an Eiger2 2D detector. Rietveld analysis was done using TOPAS software to obtain lattice parameters and crystallite sizes. Electrical conductivity and charge carrier concentration at RT were measured with a commercial HMS 5500 Hall effect measurement system (Ecopia). Seebeck measurements were done with a lab-made system in the in-plane direction at RT. For Seebeck and power factor measurements with varying temperature a commercial Linseis LSR-3 system was used.

The thermal conductivity was measured in the out-of-plane direction using the time-domain thermo-reflectance (TDTR) method at various temperatures, utilizing the Front/Front configuration. The measurements were performed with a PicoTR system (PicoTherm), employing a pump and probe laser with wavelengths of 1550 nm and 750 nm, respectively. Both lasers feature a pulse duration of 0.5 ps, and the laser pulses were applied to the film (after depositing a thin Pt layer on top of it) within a time interval of 50 ns. From these measurements, the thermal diffusivity is obtained, and from that and knowing the different parameters of the film and substrate (density of 6.584 g/cm$^3$ and heat capacity of 527 J/Kg·K), the thermal conductivity can be calculated. These measurements were conducted at the Research Institute for Electronic Science (RIES) at Hokkaido University in Japan. Details of the PicoTR system has been described elsewhere[41–43]. For estimating the lattice thermal conductivity, the electrical contribution of the lattice thermal conductivity was calculated by the Wiedemann-Franz law[44] using a Lorenz number of 2.44·10$^{-8}$ V$^2$/K$^2$ and subtracted to the total thermal conductivity measured by TDTR.

Morphological characterization was carried out via a FEI Verios 460, Scanning Electron Microscope (SEM). Atomic Force Microscopy (AFM) micrographs were obtained for two representative films of each series and are shown in the Supplementary Information (SI).

The band gap was determined through the optical reflectance using Pankove and Tauc approaches. The reflectance was measured in the mid-IR range (from 2.5 to 17 µm) with a Fourier Transform Infrared (FT-IR) spectrophotometer from PerkinElmer (Spectrum 3).

## Computational details

Ground-state structures were fully relaxed using the VASP package[45] and projector-augmented wave potentials[46]. Due to the band gap underestimation by the GGA functional, energies and band structure were calculated with the r$^2$SCAN functional proposed by Furness et al[47]. Core and valence electrons were selected following the standards proposed by Calderon et al.[48] The plane-wave basis set expansion had a kinetic energy cut-off of 500 eV, which is 25% above the standard value for the chosen PAW potentials, to reduce Pulay stress errors. Meshes of 432 and 1024 k-points per reciprocal atom was used to accurately describe the minimum of the potential energy surface and the ground state wavefunction, respectively. Geometry and lattice vectors were fully relaxed until forces on each atom were below $10^{-7}$ eVÅ$^{-1}$. The wave-function was considered converged when the energy difference between consecutive electronic steps was less than $10^{-9}$ eV, including an additional support grid for the evaluation of the augmentation charges to reduce noise in the forces.

To explore the V/Al antisite disorder in the Fe$_2$VAl bulk, we constructed a 2 × 2 × 2 periodic model containing 128 atoms. We simulated low antisite defect concentrations of 3.125% and 6.25%, with one and two antisite defects, respectively. All possible vacancy distributions were analyzed using the SOD package[49] and the high-throughput framework, DisorderNML[50]. The average properties of the disordered material were calculated by considering the contribution of each supercell to the partition function. The B2 and A2 phases were modelled using Special Quasirandom Structures (SQS[51]) as implemented in the ICET code[52]. SQS cells provide a good approximation of random alloys, as their cluster vectors closely resemble those of truly random alloys. To compare the electronic structure of the L2$_1$ phase and SQS structures, band structure unfolding[53] was performed using the easyunfold code[54].

Lattice thermal conductivity was calculated by combining the hiPhive[55] and ShengBTE[56] packages using the hiPhive wrapper[57]. The force constants were obtained using 20 4×4×4 supercells. Antisite disorder was modelled using the Tamura model[58]. Electronic transport properties were computed using the AMSET code[59], which solves the Boltzmann transport equation using the Onsager coefficients to predict electronic transport properties, with the wave function from a DFT calculation as the main input. Scattering rates were calculated for each temperature, doping concentration, band, and k-point, including scattering due to deformation potentials, polar optical phonons, and ionized impurities. More details on the methodology used to compute electronic transport properties can be found in previous works [60,61].

## Author contributions

J.M. Domínguez-Vázquez, A. Conca, A. Cebollada, O. Caballero-Calero and M. Martín-González designed the experiment and outlined the work plan for this research, J.M. Domínguez-Vázquez, A. Conca and A. Cebollada contributed to the films growth, J.M. Domínguez-Vázquez, A. Conca, A. Cebollada and K. Lohani participated in the X-ray diffraction measurements and analysis, O. Caballero-Calero conducted the thermoelectric properties measurements, M. A. Tenaguillo and H. Ohta carried out and coordinated the thermoreflectance thermal conductivity measurements, J. J. Plata and A. M. Márquez performed the computational simulations and subsequent analysis, and C. V. Manzano measured the optical band gap. AFM and SEM imaging was carried out by the MiNa services at IMN. M. Martín-González participated in the discussion with all the co-authors to achieve the final conclusions and secured financial support for A. Conca, M.A. Tenagillo, and J.M. Dominguez-Vazquez to carry out this work. Finally, J.M. Domínguez-Vázquez, A. Conca, O. Caballero-Calero, A. Cebollada, K. Lohani, J. J. Plata, C. V. Manzano and M. Martín-González contributed to drafting of the manuscript.

## Conflicts of interest

There are no conflicts to declare.

## Acknowledgements

The authors would also like to acknowledge the service from the MiNa Laboratory at IMN, and its funding from CM (project SpaceTec, S2013/ICE2822), MINECO (project CSIC13-4E-1794), and EU (FEDER, FSE). This work was funded by projects THERMHEUS grant TED2021-130874B-I00 funded by MICIU/AEI/10.13039/501100011033 and by the "European Union NextGenerationEU/PRTR" and ERC Adv. POWERbyU grant agreements ID: 101052603 Founded by European Research Council (ERC), grant PID2022-138063OB-I00 funded by MICIU/AEI/10.13039/501100011033 and by FEDER, UE. We thankfully acknowledge the computer resources at Lusitania (Cenits-COMPUTAEX), Red Española de Supercomputación, RES (QHS-2023-1-0028) and Albaicín (Centro de Servicios de Informática y Redes de Comunicaciones - CSIRC, Universidad de Granada).

# SUPPLEMENTARY INFORMATION

**Thermoelectric Performance Boost by Chemical Order in Epitaxial L21 (100) and (110) Oriented undoped Fe2VAl Thin Films: An Experimental and Theoretical Study**

José María Domínguez-Vázquez[a], Olga Caballero-Calero[a], Ketan Lohani[a], José J. Plata[b], Antonio M. Marquez[b], Cristina V. Manzano[a], Miguel Ángel Tenaguillo[a], Hiromichi Ohta[c], Alfonso Cebollada[a], Andres Conca[a,*], and Marisol Martín-González[a].

[a]Instituto de Micro y Nanotecnología, IMN-CNM, CSIC (CEI UAM+CSIC), Isaac Newton 8, E-28760 Tres Cantos, Madrid, Spain
[b]Dpto de Química Física, Facultad de Química, Universidad de Sevilla, Sevilla (Spain).
[c]Research Institute for Electronic Science, Hokkaido University, N20W10, Kita, Sapporo, 001–0020, Japan.
*Corresponding author: andres.conca@csic.es

The complete X-Ray characterization of $Fe_2VAl$ films is portrayed in SI 1, SI 2, SI 3, SI 4, SI 5 and SI 6. In SI 1 and SI 4 the specular measurements (ψ=0°) are portrayed for samples deposited over $Al_2O_3$ and MgO, respectively. These measurements are presented for the whole array of deposition temperatures ($T_{dep}$). In these figures is more visible the single-oriented nature of films deposited over MgO and polycrystalline to single-oriented transition of the ones deposited over $Al_2O_3$, as it was mentioned in the main text. MgO substrate films show the (2 0 0) X-Ray diffraction spot, meaning this that the B2 phase is present on all samples, while $Al_2O_3$ substrate ones show this feature for samples deposited at 750 °C or higher. The (1 1 1) diffraction peak, associated with $L2_1$ phase, is present for $Fe_2VAl$ films deposited either on $Al_2O_3$ or MgO at $T_{dep}$ of 850 °C or higher, as it is depicted in SI 3 and SI 6.

Another structural factor that influences thermal and transport behavior of thin films is their in-plane registry with the substrate and crystallite dispersion. We delve into it by performing ϕ–scans of asymmetric diffraction reflections of both layer and substrate (Schematic of the measurement configuration is shown in the inset to the SI 7 (a)). In SI 7 b) and c), we show ϕ–scans for the $Fe_2VAl$ (2 2 0) and (4 0 0) peaks on both substates deposited at 950 °C. For the MgO substrate, the $Fe_2VAl$ (2 2 0) scan is plotted together with MgO (2 2 0), demonstrating that $Fe_2VAl$ grows epitaxially with a 45° in-plane rotation relative to the lattice axis of MgO (see schematic in SI 7 d)). This rotation corresponds to the epitaxial relation $[1\ 0\ 0]_{Fe_2VAl}//[1\ 1\ 0]_{MgO}$ and minimizes the lattice mismatch between layer and substrate. It is worth mentioning this has been previously observed in other full Heusler systems grown on MgO with similar lattice constant [1]. Lattice parameter of MgO is 0.4213 nm, while the nominal lattice parameter of $Fe_2VAl$ is 0.5763 nm. Therefore, at 45°, the calculated lattice mismatch is $\frac{(a_{Fe_2VAl} - \sqrt{2} \cdot a_{MgO})}{(\sqrt{2} \cdot a_{MgO})}$ =3.3%. This epitaxial growth occurs for deposition temperatures of 350 °C or higher.

For the $Al_2O_3$ substrate, the ϕ–scan of $Fe_2VAl$ (4 0 0) peak is shown together with the $Al_2O_3$ (1 1 0) in SI 7 c) for the film grown at 950 °C. Here, a two-fold rotation symmetry is expected for both reflections. However, a more complex behavior is observed due to a collection of peaks originating from the layer, in addition to those from the substrate. The schematic in SI 7 d) allows identifying the different relative in-plane orientations of the different film's crystallites grown on top of the substrate. As illustrated, there is a major orientation, a secondary orientation, and one minor orientation, identified by the blue, green, and yellow arrows, respectively. The epitaxial ratio corresponding to major and secondary orientation is $[1\,1\,0]_{Fe_2VAl}//[0\,0\,1]_{Al_2O_3}$ and $[1\,0\,0]_{Fe_2VAl}//[0\,0\,1]_{Al_2O_3}$, respectively. For the preferential orientation case (blue arrows), the $Al_2O_3$ lattice parameter relevant for this orientation is $c(Al_2O_3) = 1.2993$ nm. This corresponds to a mismatch of $(\frac{2}{3}c(Al_2O_3) - \sqrt{2} \cdot a(Fe_2VAl))/(\frac{2}{3}c(Al_2O_3)) = 5.9\%$. On the other hand, for the secondary orientation (green arrows), the relevant lattice parameter is $a(Al_2O_3) = 0.4760$ nm, resulting in a mismatch of $((1.5 \cdot \frac{a(Al_2O_3)}{cos(30)} - \sqrt{2} \cdot a(Fe_2VAl))/(1.5 \cdot \frac{a(Al_2O_3)}{cos(30)}) = 14.4\%$. Thus, the preferred orientation of the lattice can be explained by the smaller mismatch with respect to the substrate. Every lattice parameter and correspondent mismatch is summarized in Table 1.

|  | **Subs. Lattice parameters** | **Lattice mismatch** |
| --- | --- | --- |
| a (Fe$_2$VAl) =0.5763 nm | a (MgO) = 0.4213 nm | 3.3% |
|  | a (Al$_2$O$_3$) =0.4760 nm<br>c (Al$_2$O$_3$) = 1.2993 nm | 14.4%<br>5.9% |

Table 1: Lattice parameters of Fe$_2$VAl and substrates and the obtained lattice mismatch for each orientation.

This behavior has previously been reported for another full Heusler alloy with a similar lattice constant but different composition, Co$_2$Cr$_{0.6}$Fe$_{0.4}$Al, pointing to a general trend with these materials and substrates [2,3]. Films deposited on Al$_2$O$_3$, for temperatures below 850 °C yield featureless, non-zero signal, ϕ-scans, indicating a complete in-plane disorder although they are still out-of-plane single-oriented for $T_{dep}$ above 800 °C.

In SI 8 a series of ϕ-scans of the (2 2 0) diffraction peak at several $T_{dep}$ of Fe$_2$VAl thin films is shown, a four-fold rotation is observed for all samples, proving epitaxial growth for the whole series. On the other hand, in SI 9, a plot of the FWHM of one of the ϕ-scan peaks is portrayed. In this figure one can see the decay of the FWHM when higher $T_{dep}$ are applied, which implies a higher in-plane order of the crystallites.

For films deposited over Al$_2$O$_3$ SI 10 shows ϕ-scans of the (4 0 0) diffraction peak at T$_{dep}$ from 800 °C to 950 °C, for lower $T_{dep}$ this plot is featureless. For $T_{dep}$ of 850 and 900 °C two preferential orientations are observed, explained in the main text, while at 950 °C the orientation (peaks at ϕ=90° and 270°) shows higher intensity.

Furthermore, to complete the information about the crystal characteristics and regain insight on lattice distortions present in the films, we have calculated the lattice interplanar distances. In SI 11, we show the dependence of the out-of-plane interplanar distances on $T_{dep}$, measured and projected from specular ($\psi=0°$) and off-specular ($\psi=45°$) XRD measurements, for films grown on MgO and Al$_2$O$_3$, respectively. This plot allows to distinguish whether the crystal structure presents a deformation from an expected cubic structure. For a perfectly cubic crystal structure, the interplanar distance of each (*h, k, l*) family of planes is defined by $d_{h,k,l}=\frac{a}{\sqrt{h^2+k^2+l^2}}$, and $a=d_{100}=4\cdot d_{400}=2\cdot d_{200}=\sqrt{8}\cdot d_{220}$, thus a difference between the last three quantities, $4\cdot d_{400}$ or $2\cdot d_{200}$ and $\sqrt{8}\cdot d_{220}$ is a proof of non-isotropic distortion of the lattice. For comparison reasons, in SI 11, the interplanar distances are plotted together with the reported bulk lattice parameter obtained from[4]. SI 11 b), depicts in-plane interplanar distances for the three highest $T_{dep}$ due to the absence of (*h* 0 0) peaks for lower temperature cases. For both substrates, the increasing $T_{dep}$ results in a decrease in the lattice parameter, reaching values close to the ones reported on bulk samples, particularly for films deposited on Al$_2$O$_3$, where the difference is less than $2\cdot10^{-3}$nm. Lattice parameter and crystallite size extracted from Rietveld analysis, and their dependence on $T_{dep}$, are shown in the SI. In brief, the crystallite size (in the order of 20 nm) gradually increases with increasing growth temperature, as expected in these kinds of thin films.

To complete the information about the crystal characteristics of these films, in SI 12 we show the growth temperature dependence out-of-plane crystallite size for both layer orientations obtained from Rietveld analysis of the symmetrical and asymmetrical XRD θ/2θ scans. As it can be seen, the out-of-plane crystallite size gradually increase with increasing $T_{dep}$, which is the usual behavior in thin film growth, films grown on MgO appear to have a tendency to increase faster with $T_{dep}$ than the ones deposited over Al$_2$O$_3$.

In addition to electronic transport and thermoelectric properties, charge carrier concentration was measured in all samples and is included in SI 13, although the thin nature of the films affects the quality of the measurements, making it impossible to differentiate the concentration between samples. However, all the measured values are in the range of $10^{20}$cm$^{-3}$. Scanning electron microscope (SEM) images of all samples are shown in SI 14 and SI 15, the former showing Fe$_2$VAl deposited over Al$_2$O$_3$ and the latter over MgO. AFM measurements were performed on samples deposited at 550 and 850 °C over the two substrates and are shown in SI 16 and SI 17, where it is visible on the profiles at the right of the images that the surface roughness increases significantly with T$_{dep}$. Also, extracted values of RMS are shown in Table 2.

The effect of chemical disorder on thermal transport is done via a comparison of measured and calculated thermal transport properties, in SI 18 the calculated lattice thermal conductivity is plotted along with the estimated values obtained on two measured samples. This estimation was made subtracting the electronic contribution, obtained with the Wiedemann-Franz law ($\kappa_e=\sigma$LT)[5] using a Lorenz number of *L=2.4·10$^{-8}$ V$^2$K$^{-2}$*, to the total thermal conductivity measured with TDTR.

The thermoelectric properties dependence on carrier concentration and crystallite size is shown in SI 19 as a form of heat maps.

The optical band gap measurement done with Tauc plot is shown in SI 20, this measurement gives the same result as the Pankove plot shown in the main text (E$_g$=0.19±0.05 eV).

| Substrate | $T_{dep}$ (°C) | RMS (nm) |
|---|---|---|
| MgO | 550 | 0.57 |
| | 850 | 3.95 |
| $Al_2O_3$ | 550 | 1.17 |
| | 850 | 3.98 |

Table 2: Root Mean Square (RMS) roughness values obtained for samples deposited at 550 and 850°C over $Al_2O_3$ and MgO.

# XRD

## Al₂O₃ substrate:

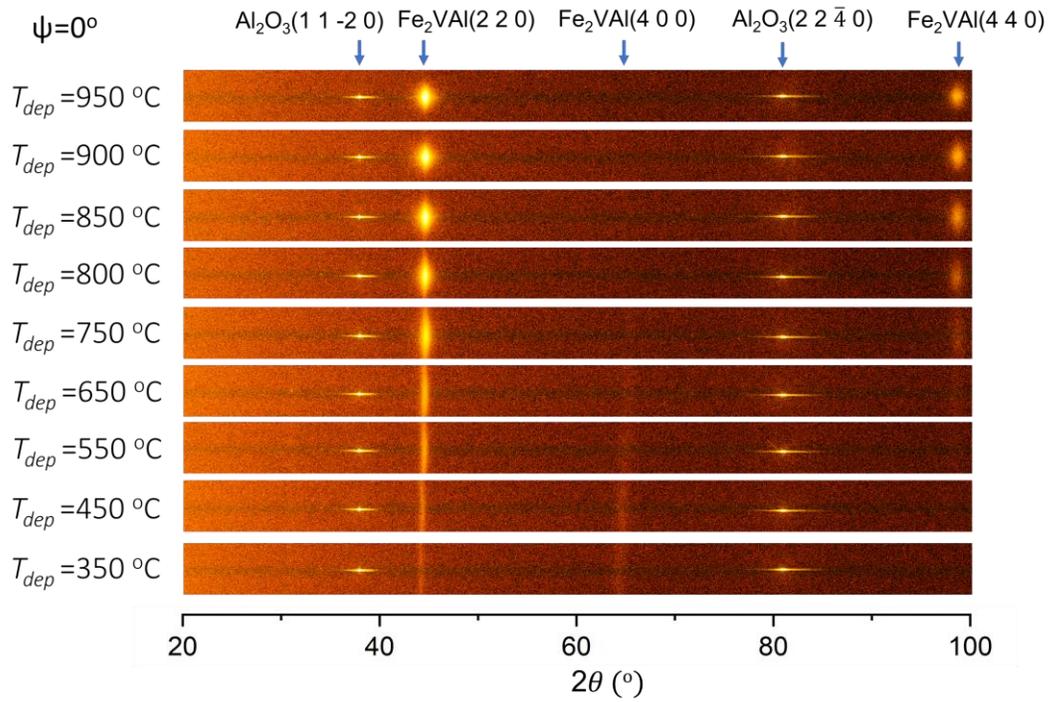

SI 1: specular (ψ=0º) X-Ray $\theta/2\theta$ vs $\gamma$ 2D maps for Fe$_2$VAl films at all deposition temperatures (T$_{dep}$) deposited over Al$_2$O$_3$ substrate.

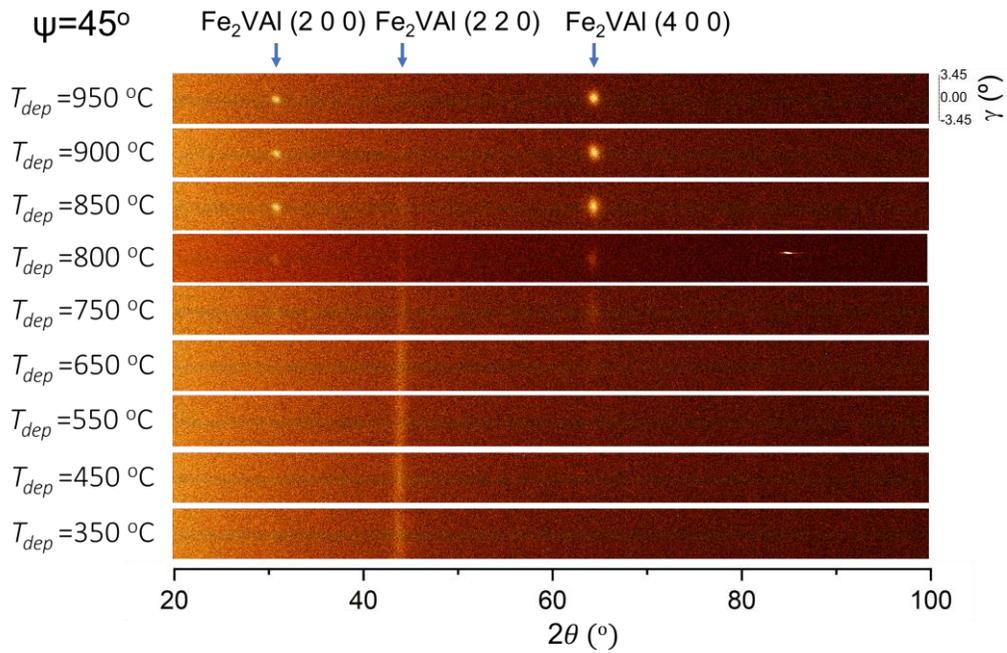

SI 2: Off-specular (ψ=45º) X-Ray $\theta/2\theta$ vs $\gamma$ 2D maps for Fe$_2$VAl films at all deposition temperatures (T$_{dep}$) deposited over Al$_2$O$_3$ substrate.

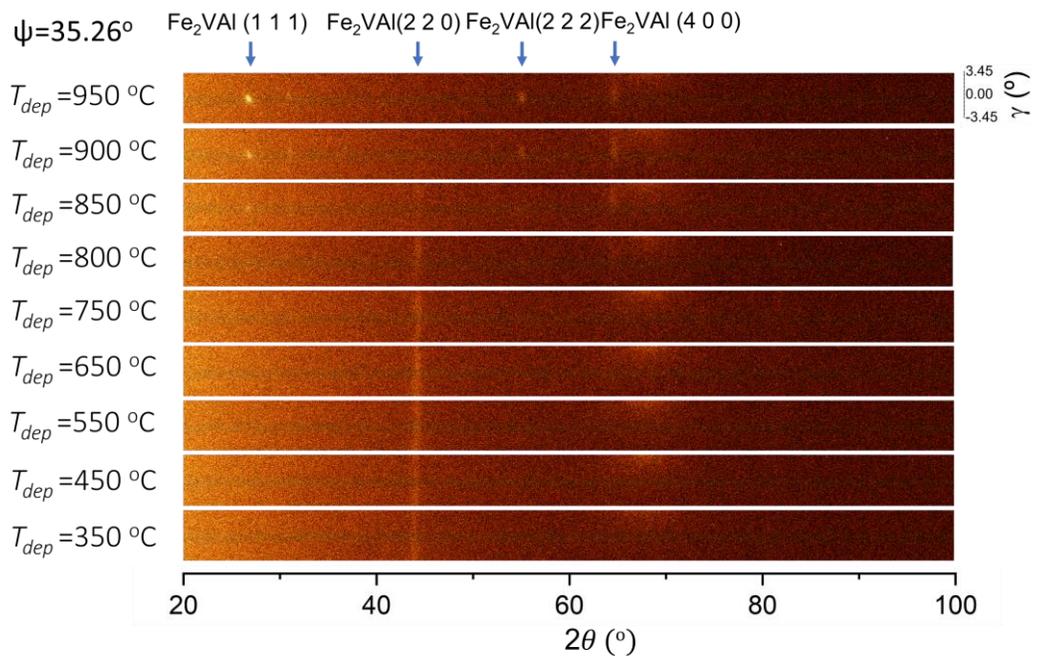

SI 3: Off-specular (ψ=35.26°) X-Ray $\theta/2\theta$ vs $\gamma$ 2D maps for Fe$_2$VAl films at all deposition temperatures (T$_{dep}$) deposited over Al$_2$O$_3$ substrate.

**MgO substrate:**

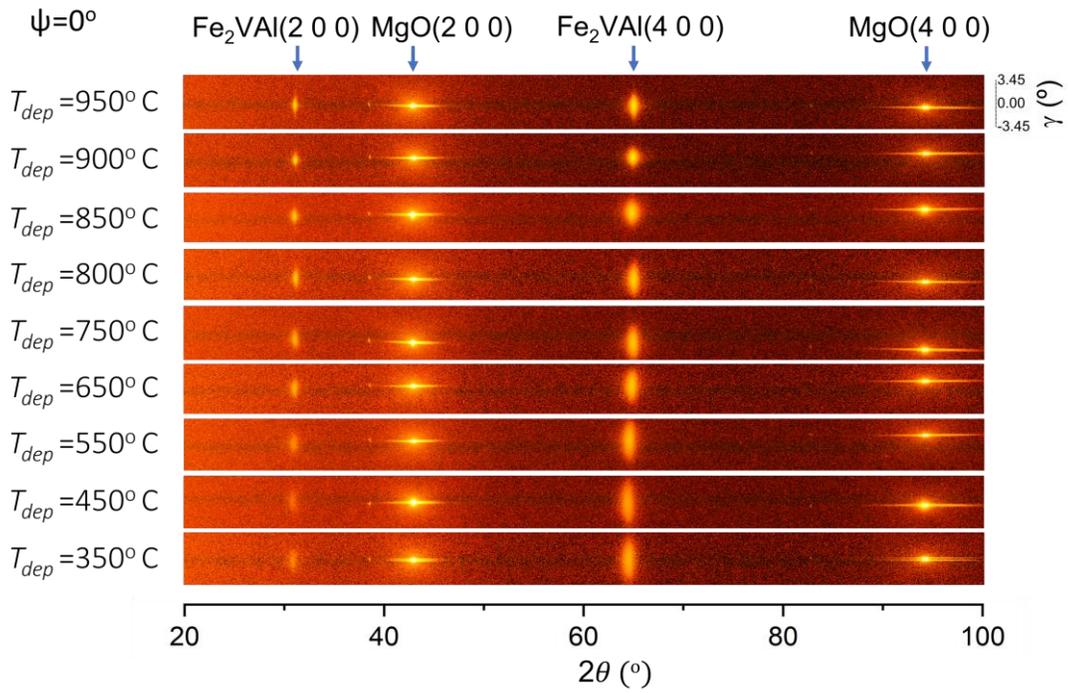

SI 4: specular (ψ=0º) X-Ray $\theta/2\theta$ vs $\gamma$ 2D maps for Fe$_2$VAl films at all deposition temperatures (T$_{dep}$) deposited over MgO substrate.

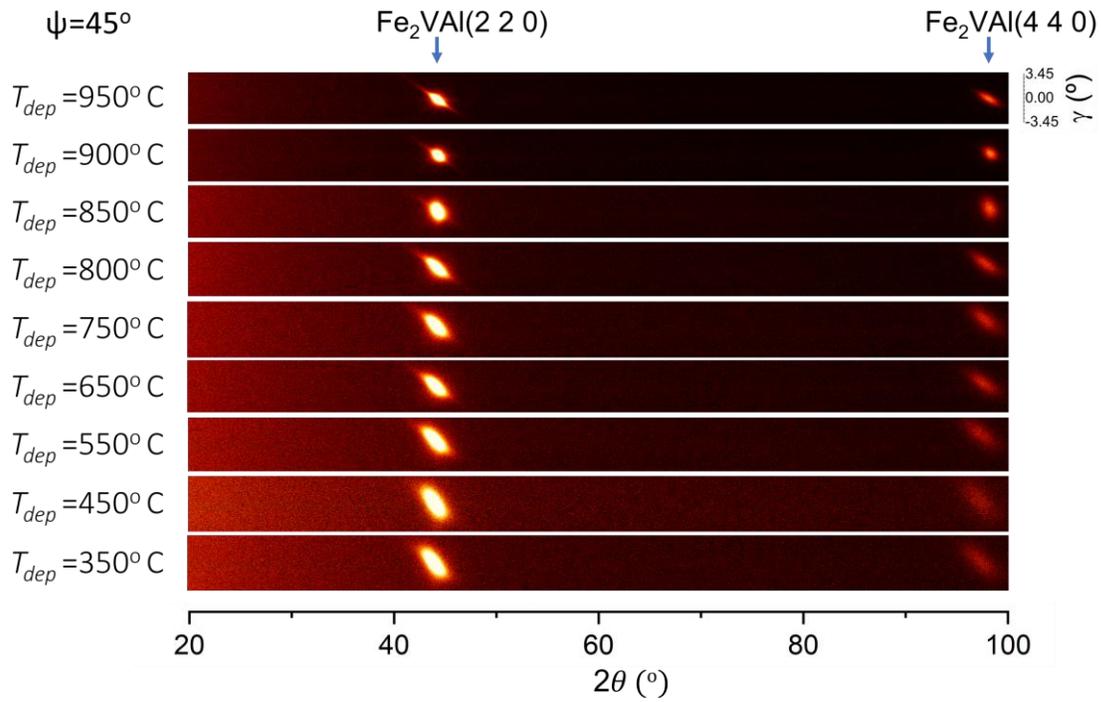

SI 5: Off-specular (ψ=45º) X-Ray $\theta/2\theta$ vs $\gamma$ 2D maps for Fe$_2$VAl films at all deposition temperatures (T$_{dep}$) deposited over MgO substrate.

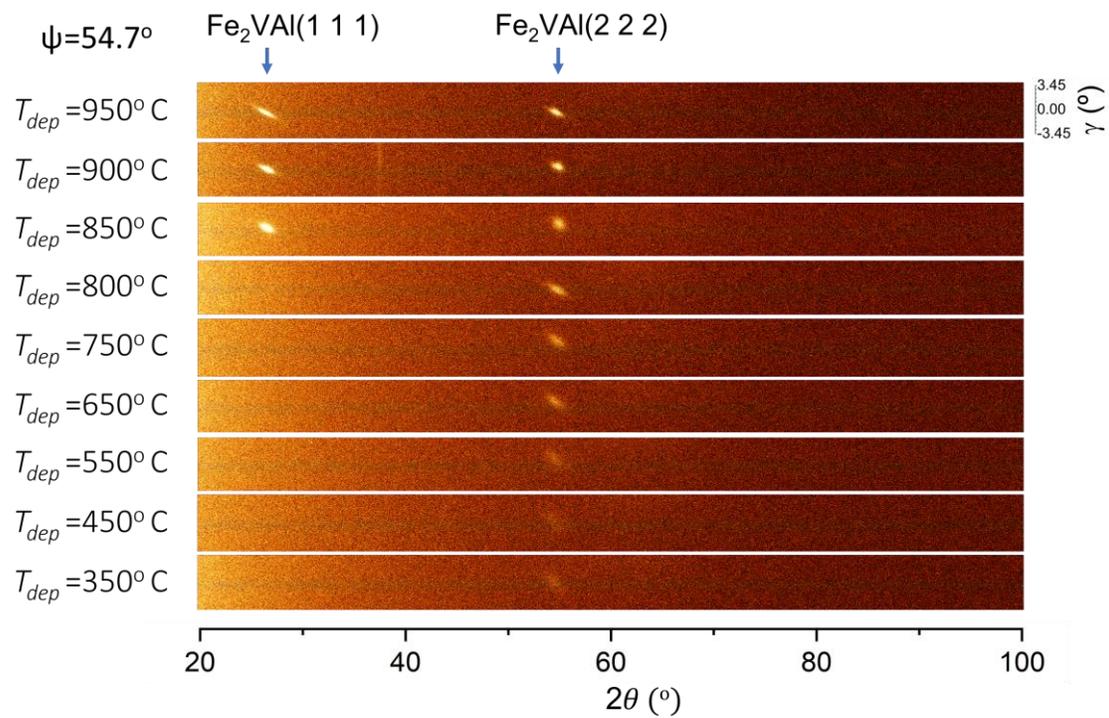

SI 6: Off-specular (ψ=54.7°) X-Ray $\theta/2\theta$ vs $\gamma$ 2D maps for Fe$_2$VAl films at all deposition temperatures (T$_{dep}$) deposited over MgO substrate.

# Phi scans, interplanar distances and crystallite size

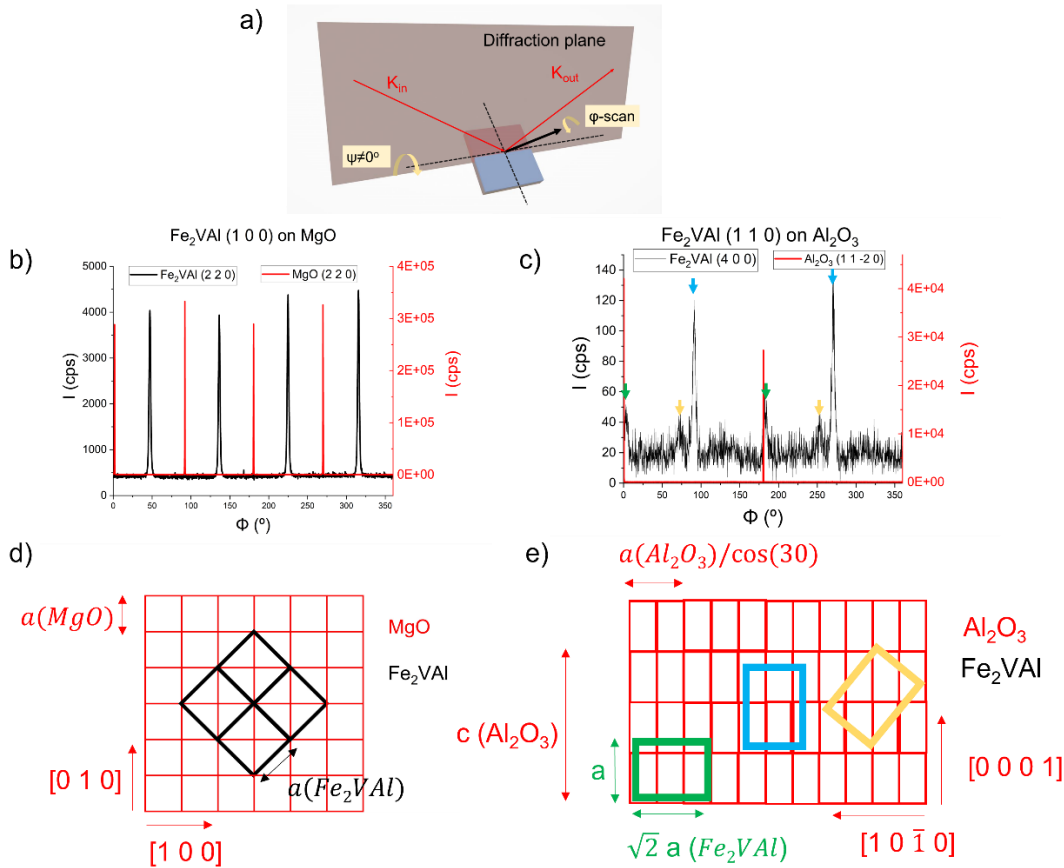

SI 7: ϕ–scans of asymmetrical peaks of Fe$_2$VAl thin films. a) schematic diagram of the ϕ-scan measurement geometry. For films grown at 850 °C on MgO b) the (2 2 0) peak is plotted and for the ones grown at 950 °C on Al$_2$O$_3$ c) the (4 0 0). d) and e) are schematic of the in-plane texture of the films based on the ϕ–scans showed on b) and c), respectively.

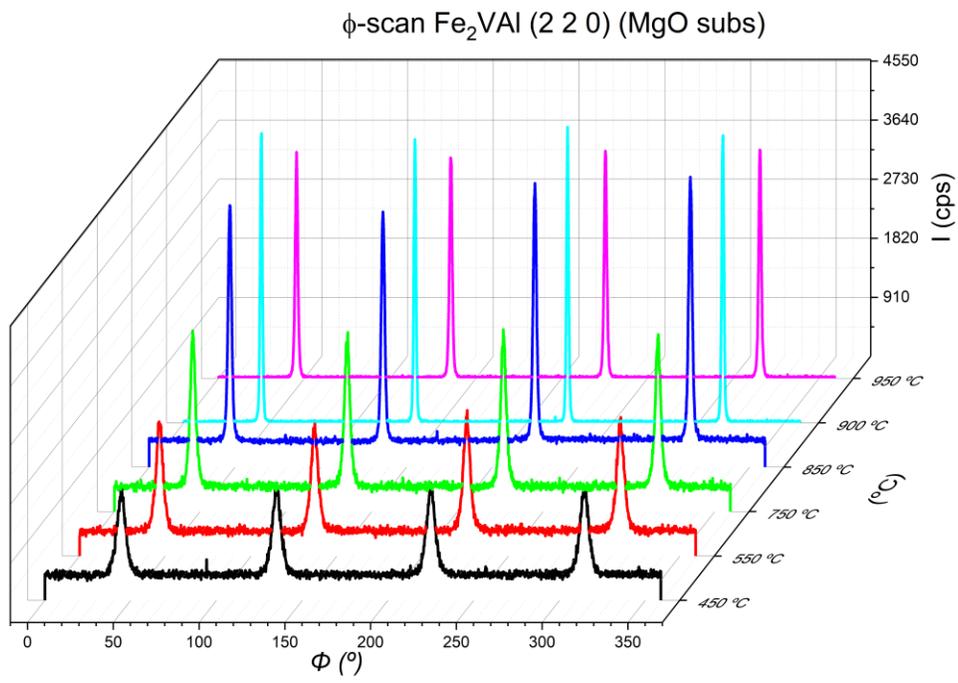

SI 8: ϕ-scan of (2 2 0) diffraction peak of Fe₂VAl films deposited over MgO for various T$_{dep}$.

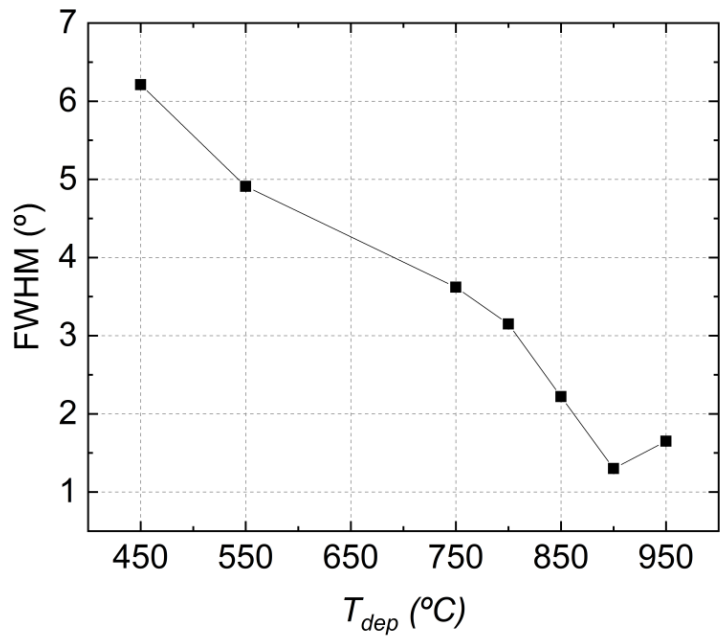

SI 9: Full Width Half Maximum (FWHM) of ϕ-scan peaks of Fe₂VAl films deposited over MgO for various T$_{dep}$.

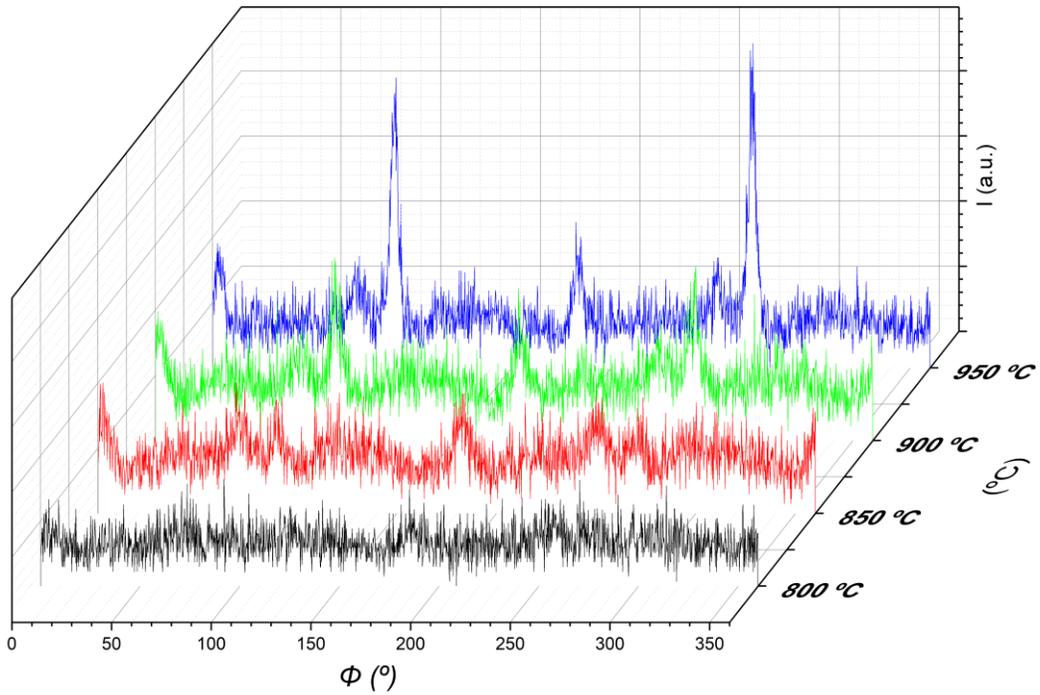

SI 10: ϕ-scan of (4 0 0) diffraction peak of Fe$_2$VAl films deposited over Al$_2$O$_3$ for various T$_{dep}$.

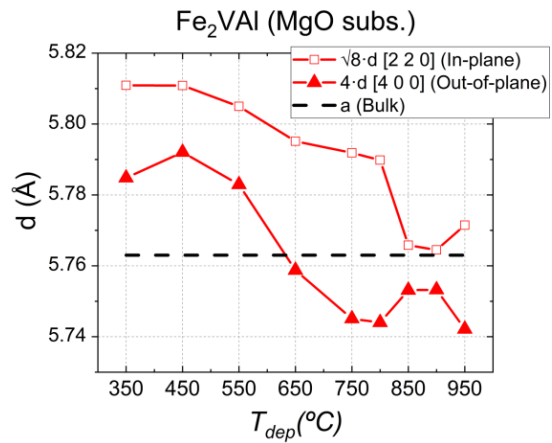
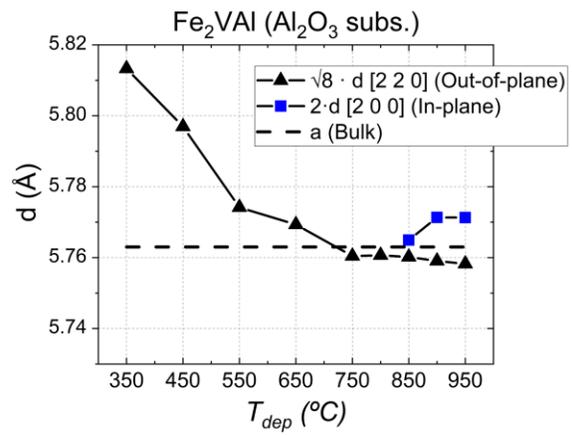

SI 11: Measured out-of-plane (ψ=0°, triangle markers) and in-plane (ψ=45°, square markers) interplanar distances of Fe$_2$VAl deposited on MgO a), and Al$_2$O$_3$ b). The interplanar distances were compared to estimate the deviation of the lattice from a cubic crystal structure, as in this type of crystal structure . The measured distances are compared with the bulk lattice parameter of Fe$_2$VAl from [4].

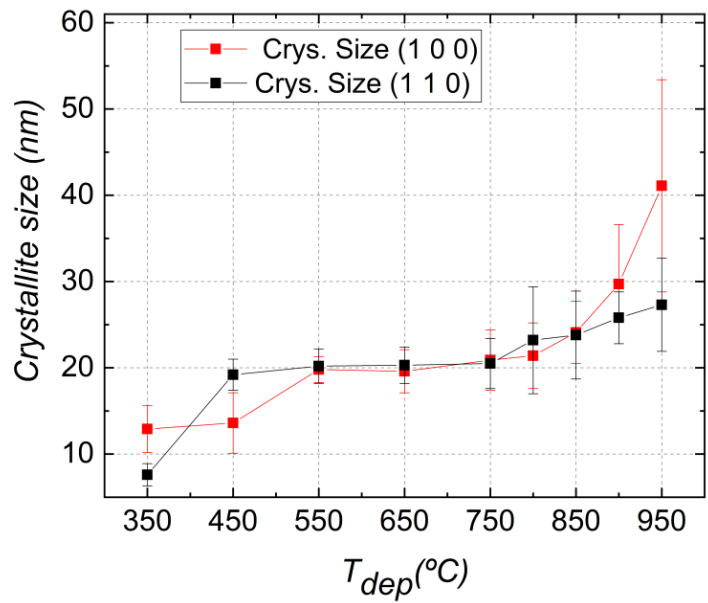

SI 12: Out-of-plane crystallite size obtained from Rietveld analysis of the specular (ψ=0°) $\theta/2\theta$ measurements of $Fe_2VAl$ (1 0 0) and (1 1 0) films.

## Carrier concentration

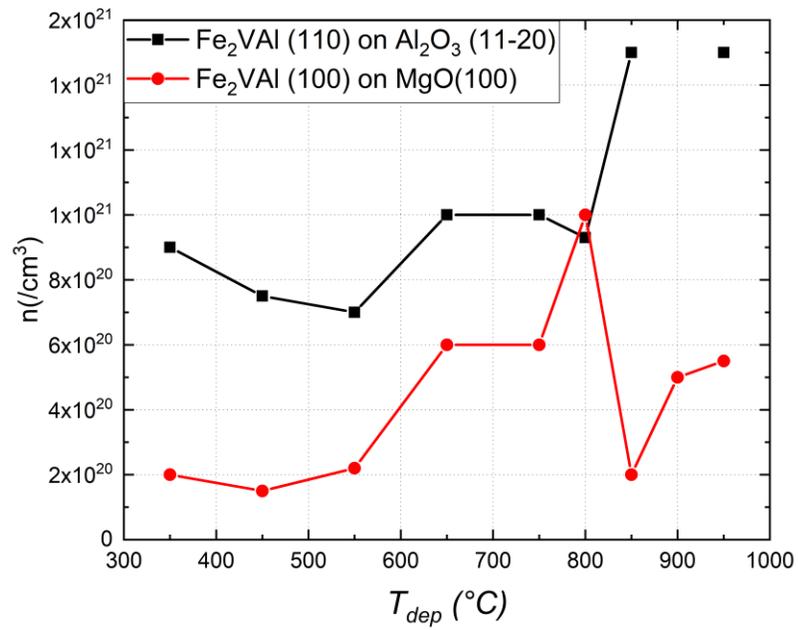

SI 13: Measured Carrier concentration of the Fe$_2$VAl (1 1 0) and (1 0 0) films deposited at different deposition temperatures.

## SEM

## Al$_2$O$_3$ substrate:

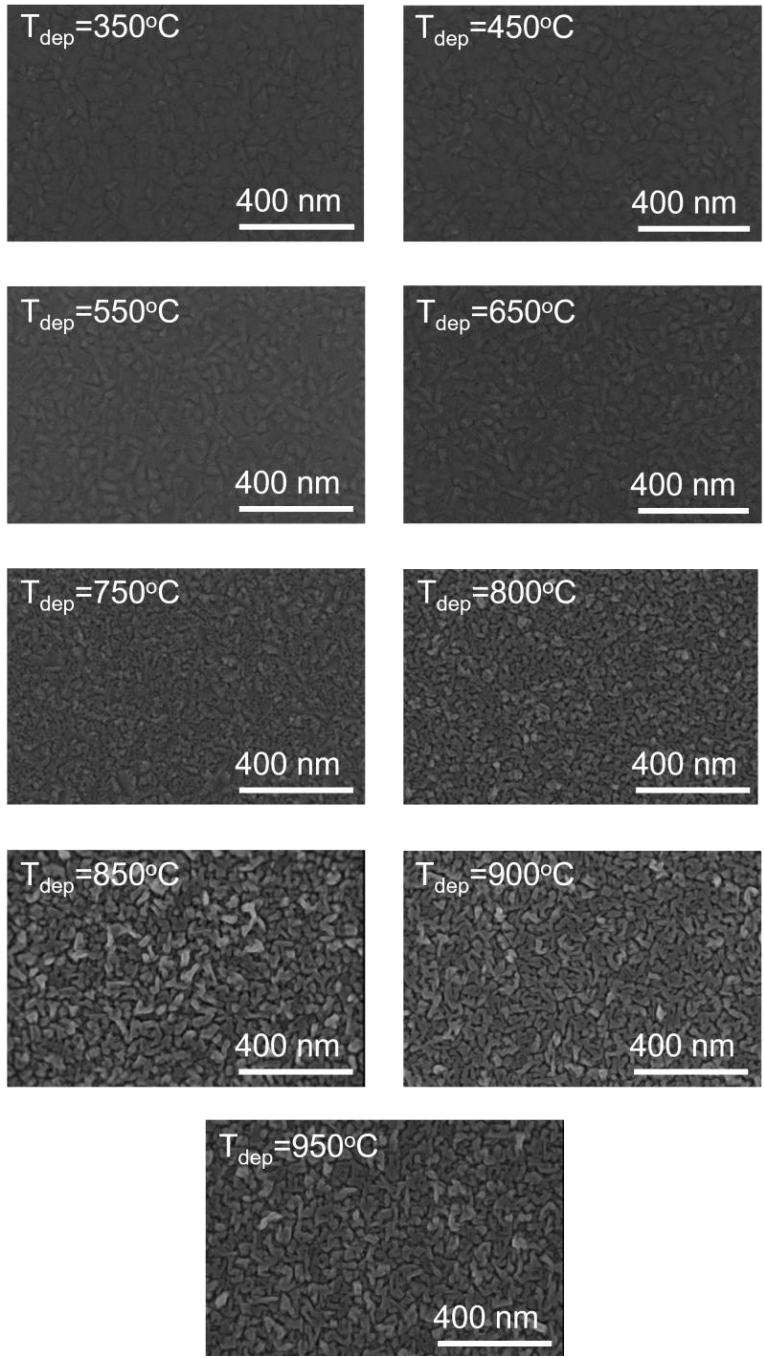

SI 14: SEM images of Fe$_2$VAl films deposited over Al$_2$O$_3$ at all T$_{dep}$.

## MgO substrate:

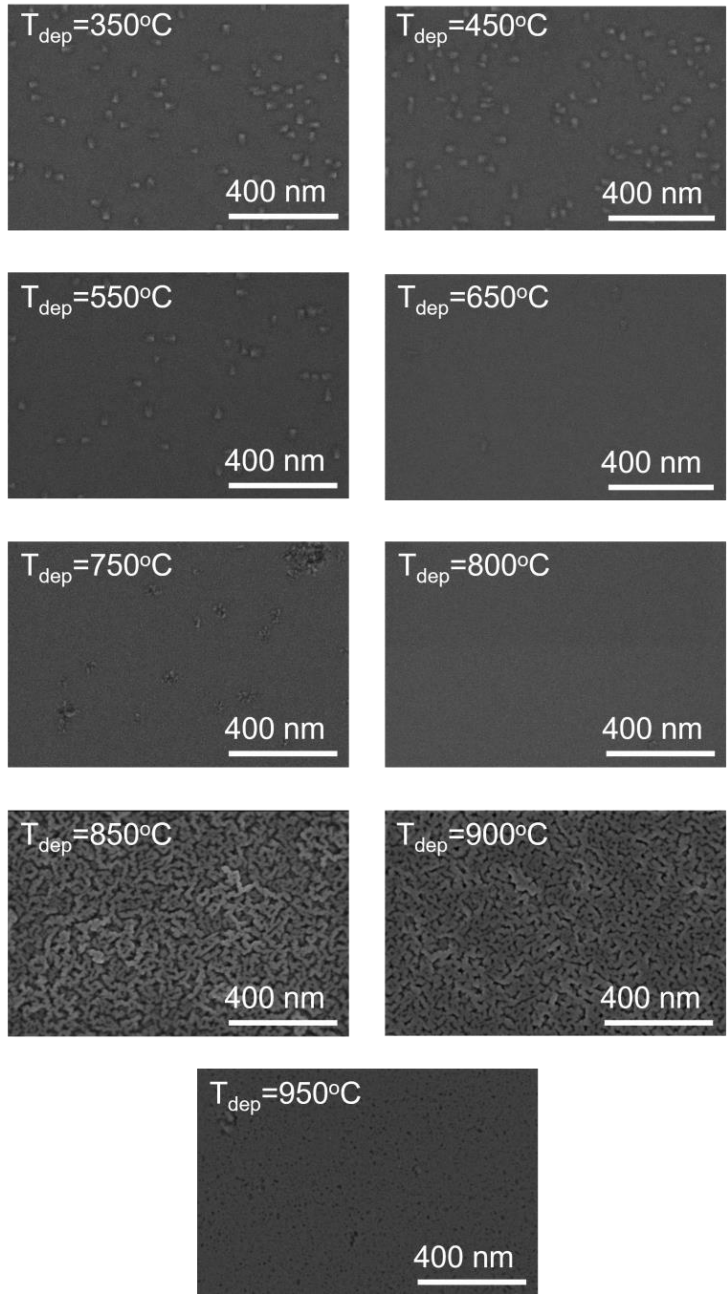

SI 15: SEM images of $Fe_2VAl$ films deposited over MgO at all $T_{dep}$.

**AFM**

## MgO subs.

$T_{dep}=850°C$

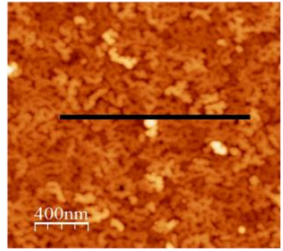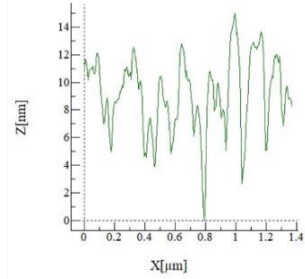

$T_{dep}=550°C$

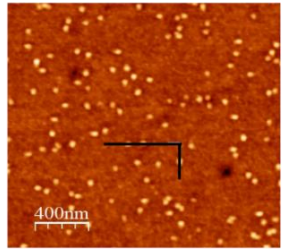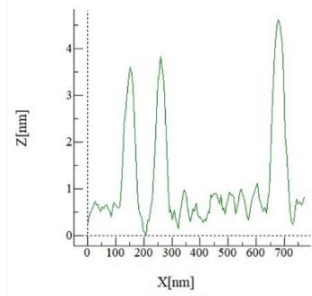

SI 16: AFM images and profile for samples deposited at 550 and 850 °C over MgO.

## $Al_2O_3$ subs.

$T_{dep}=850°C$

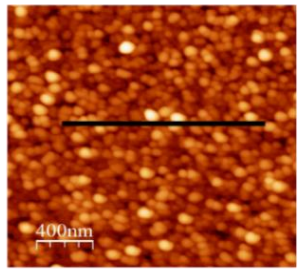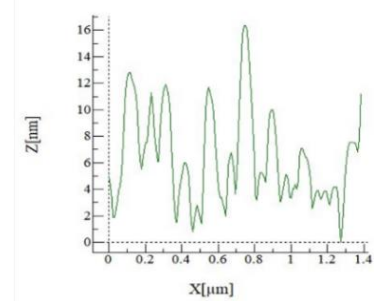

$T_{dep}=550°C$

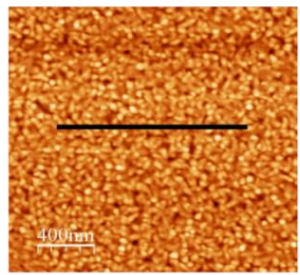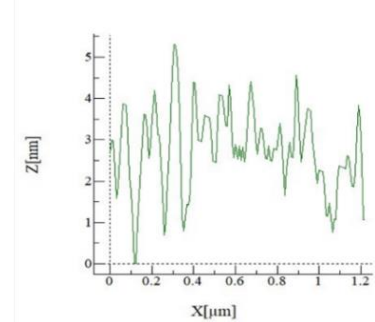

SI 17: AFM images and profile for samples deposited at 550 and 850 °C over $Al_2O_3$.

## Additional simulations

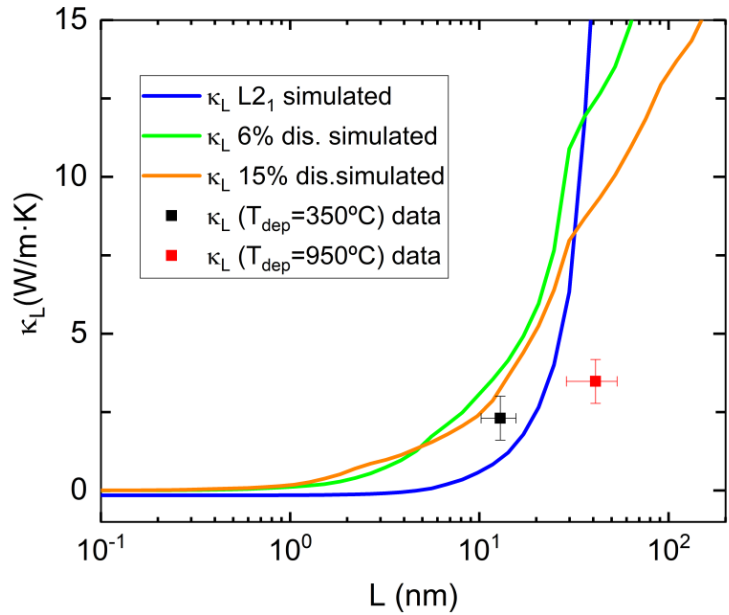

SI 18: Comparison between simulations and measurements of the lattice contribution to thermal conductivity $\kappa_l$ as a function of the crystallite size.

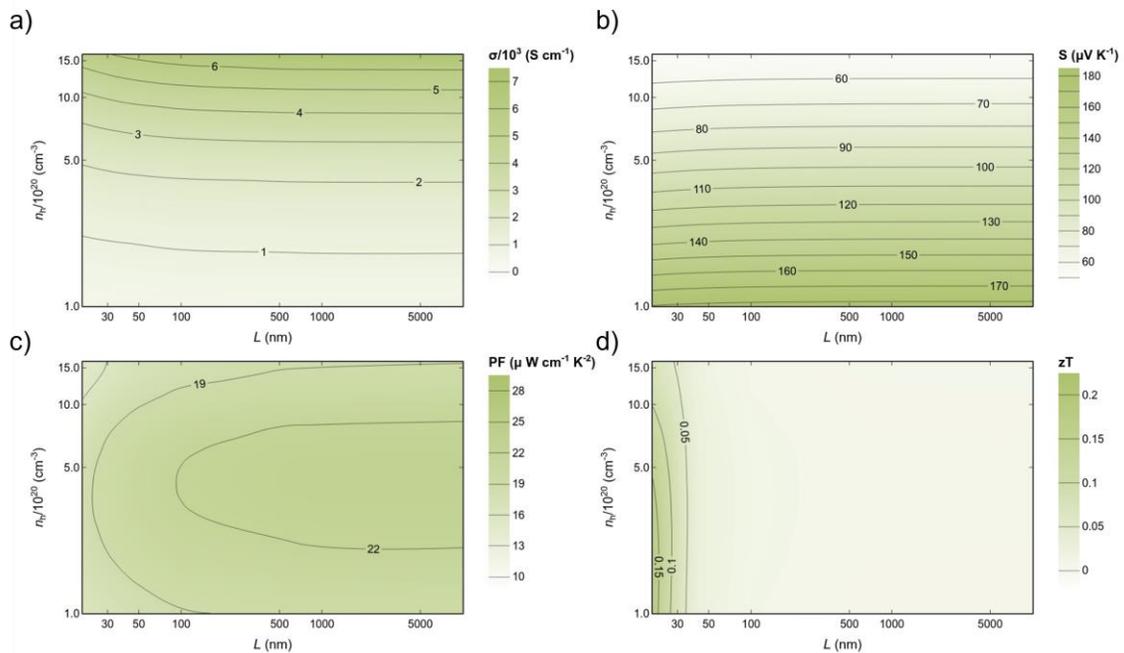

SI 19: Heat maps of a) Electrical conductivity ($\sigma$), b) Seebeck coefficient (S), c) Power Factor (PF), and d) thermoelectric figure of merit (zT) dependence on carrier concentration (n) and grain size (L) at 300 K for $L2_1$ $Fe_2VAl$ phase.

## Optical band gap

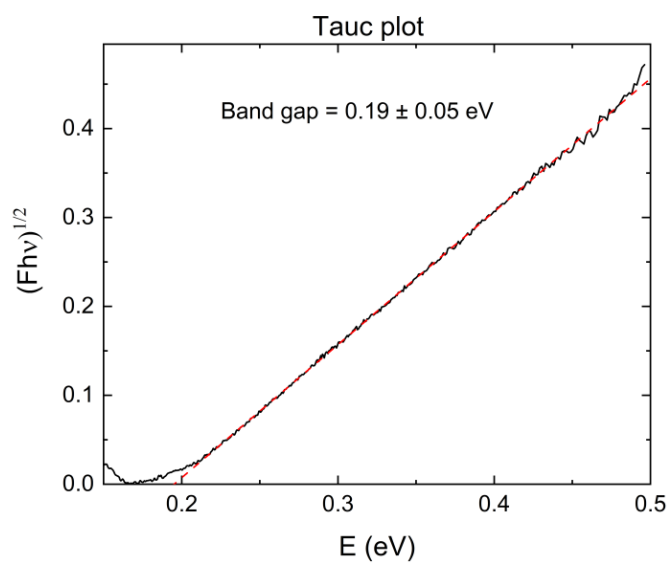

SI 20: Optical measurement of the band gap of $Fe_2VAl$ via Tauc plot. The estimation gives a result of $E_g$=0.19±0.05 eV.